\documentclass[conference]{IEEEtran}
\usepackage{blindtext, graphicx}
\usepackage[utf8]{inputenc}
\usepackage[english]{babel}
\usepackage{amsmath}
\usepackage{algorithm}
\usepackage{algpseudocode}
\usepackage[skip=4pt]{caption}
\usepackage{multirow}
\usepackage{array}
\usepackage{booktabs}
\usepackage{soul}
\usepackage{graphicx}
\usepackage{subcaption}

\usepackage[usenames, dvipsnames]{color}
\ifCLASSINFOpdf
\else
\fi
\renewcommand{\baselinestretch}{0.97}
\begin{document}
%

\title{Leader-follower based Coalition Formation in Large-scale UAV Networks, A Quantum Evolutionary Approach}

\author{
   \IEEEauthorblockN{
     Sajad Mousavi\IEEEauthorrefmark{1},
     Fatemeh Afghah\IEEEauthorrefmark{1},
    Jonathan D. Ashdown\IEEEauthorrefmark{2} 
    and
    Kurt Turck\IEEEauthorrefmark{2}}
   \IEEEauthorblockA{
     \IEEEauthorrefmark{1}School of Informatics, Computing and Cyber Systems,
    Northern Arizona University, Flagstaff, AZ, United States\\
     Email: \{sajadmousavi, fatemeh.afghah\}@nau.edu}
   \IEEEauthorblockA{
   \IEEEauthorrefmark{2}Air Force Research Laboratory,
   Rome, NY, United States\\
     Email: \{jonathan.ashdown, kurt.turck\}@us.af.mil}
 }


%


\maketitle

\begin{abstract}
The problem of decentralized multiple Point of Interests (PoIs) detection and associated task completion in an unknown environment with multiple resource-constrained and self-interested Unmanned Aerial Vehicles (UAVs) is studied. The UAVs form several coalitions to efficiently complete the compound tasks which are impossible to be performed individually.
The objectives of such coalition formation are to firstly minimize resource consumption in completing the encountered tasks on time, secondly to enhance the reliability of the coalitions, and lastly in segregating the most trusted UAVs amid the self interested of them. As many previous publications have merely focused upon minimizing costs, this study considers a multi-objective optimization coalition formation problem that considers the three aforementioned objectives. In doing so, a leader-follower-inspired coalition formation algorithm amalgamating the three objectives to address the problem of the computational complexity of coalition formation in large-scale UAV networks is proposed. This algorithm attempts to form the coalitions with minimally exceeding the required resources for the encountered tasks while maximizing the number of completed tasks. The proposed algorithm is based on Quantum Evolutionary Algorithms (QEA) which are a combination of quantum computing and evolutionary algorithms. Results from simulations show that the proposed algorithm significantly outperforms the existing coalition formation algorithms such as merge-and-split and a famous multi-objective genetic algorithm called NSGA-II \footnote{DISTRIBUTION A. Approved for public release: distribution unlimited. Case
Number: 88ABW-2018-0096 . Dated 10 Jan 2018.
}.
\end{abstract}

\begin{IEEEkeywords}
Unmanned aerial vehicles, coalition formation, mission completion, evolutionary algorithms.
\end{IEEEkeywords}

%
\IEEEpeerreviewmaketitle

\section{Introduction}
A single agent system is often unable to perform complex tasks considering the limited individual capabilities of such an agent. Therefore, cooperative multi-agent systems (MASs) can offer a practical solution to this problem by ensembling a complementary set of different capabilities/resources from several agents. However, one of the key challenges in such cooperative MASs is forming optimal sub-groups of agents (i.e., coalition formation) in order to efficiently perform the existing tasks, especially in a distributed case where no central controller is available. That being said, the coalition formation problem concerns how different coalitions can be formed considering the tasks' requirements and the agents' capabilities so much so that the collective goals of the tasks are reached in the most effective manner.

A considerable amount of research has been recently carried out in solving coalition formation problems. This has spawned several classic methods to form stable coalitions that follow common stability concepts based on Core, Shapley value, Bargaining Set, and Kernel \cite{sandholm1999distributed,ketchpel1993coalition,contreras1998multi}. However, achieving such stability concepts often mandates high computational complexity. Many researchers have attempted to deal with the problem of coalition formation in multi-agent systems by applying various approaches including genetic algorithms \cite{yang2007coalition}, dynamic programing methods \cite{cruz2013optimizing}, graph theory \cite{bistaffa2014anytime,sless2014forming}, iterative processes \cite{janovsky2016multi}, cooperative Multi-Agent Reinforcement Learning (MARL) \cite{ghazanfari2014enhancing}, and temporal-spatial abstraction MARL \cite{ghazanfari2017autonomous,ghazanfari2016extracting,mousavi2014automatic}. The majority of these reported studies have focused mainly in reducing costs and only a few of them have focused upon multiple objectives in solving the problem. In this paper, a solution for coalition formation problem in a heterogeneous resource-constraint UAV network in which multiple objectives are considered to form the optimal coalitions is proposed.
The objectives of the coalition formation method proposed here are : minimizing the cost associated with consumption of resources of the coalitions formed, maximizing the reliability of the formed coalitions, and lastly to select the most trustworthy of UAVs among the available self-interested UAVs in the network. 

Finding the solution of such a multi-objective coalition formation problem involves an NP-hard problem. Many approaches such as mixed integer linear programming \cite{schumacher2004uav,darrah2005multiple} and dynamic network flow optimization \cite{schumacher2002task} have been utilized to provide an exact solution to this problem. However, since these approaches seek such a solution, applying them to a large-scale problem is computationally taxing. More recently, metaheuristic algorithms such as Particle Swarm Optimization (PSO) \cite{kennedy2011particle}, Ant Colony Optimization \cite{dorigo1997ant}, Genetic Algorithms (GAs) \cite{goldberg2006genetic}, and Simulated Annealing (SA) \cite{van1987simulated} have offered reasonable solutions in efficient times for a variation of multi-objective optimization problems.  Inspired by the success of evolutionary approaches, this paper presents a novel leader-follower-based coalition formation algorithm using a quantum evolutionary approach whilst considering the aforementioned objectives in addressing the problem. As an unknown, dynamic environment is assumed where no prior knowledge is available about targets or Point of Interests (PoIs) and UAVs need to gain the knowledge of environment dynamically \cite{8124890,razi2017optimal}. Some potential applications of the proposed method are search and rescue, humanitarian relief, and public safety operations in unknown remote environments or military operations in remote fields. In such environments, it can be safely assumed that ground station does not have prior information about the PoIs and their positions. To evaluate the performance of the proposed method in such an unknown environment, several scenarios with different numbers of tasks ranging from 4 to 24, and a heterogeneous network of UAVs consists of a different number of UAVs ranging from 8 to 124 were simulated. 

The rest of the paper is organized as follows: Section \ref{Sec:formulation} presents the problem statement and the formulation of the multi-UAV coalition formation as a multi-objective optimization problem. Section \ref{Sec:proposed} describes the proposed coalition formation algorithm. Section \ref{Sec:results} reports the simulation results followed by concluding remarks in section \ref{Sec:conclusion}.

\section{Problem Statement}
\label{Sec:formulation}
In this section, the decentralized task allocation problem in a network of autonomous UAVs by forming optimum coalitions are formulated. 
The possibility of selfish behavior of these self-interested UAVs are accounted for and reputation guidelines in selecting the most reliable of UAVs to participate in the formed coalitions are also defined. It is vital to note that the proposed algorithm considers minimizing the cost of coalition formation in terms of overspending the resources on particular tasks as well as enhancing the reliability of the formed coalitions as paramount.

A heterogeneous network of $N$ UAVs $\mathcal{U}=\{u_1,u_2,\ldots,u_N\}$, where each UAV, $u_i$ can carry a different set of resources compared to another is considered. $R_{u_i}=\{r_{u_i }^1,r_{u_i}^2,\ldots,r_{u_i}^{N_r}\}$ denotes the set of resources available at UAV, $u_i$ where $N_r$ is the number of possible resources in the network. It is also assumed that there exists $n$ tasks in the environment $T=\{T_1,T_2,\ldots,T_n\}$. Each task, $T_i$ requires a certain amount available resource in each of $R_{u_i}$ in order to be completed. The vector of required resources for task $i$ is defined as follows: \begin{equation}
\tau^i=\{\tau^i_1,\tau^i_2,\tau^i_3,\ldots,\tau^i_{N_r}\},
\end{equation}
where $\tau^i_j$ is the required amount of resource $j$ for task $i$. It is assumed that each task is associated to one PoI (target) and the PoIs can be located in different positions with a diverse set of resource requirements. All UAVs are able to search the unknown environment for new PoIs. The tasks are carried out by the formed coalitions $\mathcal{S}=\{S_1,S_2,\ldots,S_m\}$, where each coalition $S_i$ is responsible for one task. A large search space where the PoIs are distributed far apart from each other are considered. It can therefore be assumed that the formed coalitions are sufficiently far from one another that each UAV can only be a member of a single coalition, i.e., $S_k\cap S_l=\emptyset, \forall S_k, S_l \in \mathcal{S}$. This also means that the coalitions are non-overlapping. The capability of each coalition $S_i$ to complete its encountered task is defined as the value of coalition $v(S_i), ~(v(S_i) \in \mathcal{R})$ as described in the next section. Moreover, $cost(S_i,T_i)$ is defined as the cost of coalition $S_i$ in performing task $T_i$. The cost function for coalition $S_i$ captures the cost imposed on all UAV members of this coalition (i.e., $\sum_{u_j\in S_i} cost(S_i,u_j))$ in which their resources have been shared to accomplish task {$T_i$}. 

Another key contribution of the proposed model is considering the reliability factor in forming the optimal coalitions. While in the majority of existing techniques, it is assumed that the UAV members of formed coalitions are perfectly operational during the mission lifetime, this is obviously not a realistic assumption as the UAVs' operation can be interrupted for several reasons (e.g., exhaustion of battery or a particular resource). A practical scenario is considered where the UAVs are assumed to be either fully operational or one of their capabilities (e.g., resources) are bound to fail during the mission. Such failures in various capabilities are considered to be statistically independent. Furthermore, involving different types of UAVs in a coalition may result in different execution times of accomplishing the sub-tasks as the UAVs have a different set of capabilities. For instance, a given UAV may be able to fulfill its duty in a shorter time than another. The cost and reliability of a coalition indeed depend on these execution times where lower execution times are favored for incurring lower execution costs.
In the next section, we define and formulate these factors (i.e., cost, reliability and reputation).   

\subsection{Definition of Cost, Reliability, and Reputation}
A set of UAVs in the form of a coalition collaborate with one another to carry out the encountered task. Participation in such coalitions involves a cost of sharing and consuming the resources for the member UAVs. For a given task $T_k$ with the required resources $\tau^k=\{\tau^k_1,\tau^k_2,\tau^k_3,\ldots,\tau^k_{N_r}\}$, where the amount of resource $i$ for task $k$ is denoted by $\tau^k_i\geq 0$, the execution cost of consuming resource $j$ of UAV $i$ is denoted by $e_{ij}$, where $e_{ij}=\mu_j r_{u_i}^j \quad \forall i,j$ and $\mu_j$ is a constant coefficient in order to convert the amount of resource $j$ to a time dependent value to have the same unit as the execution time. Thus, the cost of  coalition $S_k$, $C(S_k)$ can be calculated as follows:
\begin{equation}
C(S_k)=\sum_{i=1}^{N_k}\sum_{j=1}^{N_r} {e_{ij}k_{ij}+a_{ij}},
\end{equation}
where $k_{ij}$ is the execution time if UAV $i$ carrying resource $j$ is involved in task $k$, and $a_{ij}$ is the travel time of UAV $i$ to task $j$.

Possibility of potential defects in the UAVs' resources that may result in performance failure of these UAVs during the mission is also accounted for. Considering so, the reliability of coalition $S_k$ for a give task $T_k$, denoted by $R(S_k)$ is defined as follows:
\begin{equation}
R(S_k)=\prod_{i=1}^{N_k}e^{-\sum_{j=1}^{N_r} {\lambda_{ij}k_{ij}}},
\end{equation}
where $\lambda_{ij}$ is the failure rate of resource $j$ of UAV $i$.
For simplicity, $\log_e$ transfer of $R(S_k)$ function as follows, $\ln(R(S_k)) = -\sum_{i=1}^{N_k}\sum_{j=1}^{N_r} {\lambda_{ij}k_{ij}}$ is used.
The formulation of the reliability has been inspired by works reported in \cite{han2000genetic} and \cite{yin2007multi}. Interested readers are referred to these papers for more details on the probability that a system can accomplish a particular task without failure. Similar to other cognitive agents, the UAVs are expected to be self-interested in the sense that they prefer to save their limited resources, and act selfishly by not consuming enough resources during the mission. To monitor the cooperative behavior of these UAVs, an accumulative cooperative reputation related to the amount of resources that each UAV shares during the mission is defined. 
During each mission $t$ (i.e., accomplishing an assigned task), the cooperative reputation of each UAV $i$, $\rho_i$ is updated as follows:
\begin{equation}
\rho_i^t=\bigg \{ \begin{tabular}{cc}
  $\rho_i^{t-1} + \Delta \rho_i^t$, & $\exists k|u_i \in S_k$ \\
  $\rho_i^{t-1}$, & otherwise  \\
  \end{tabular}
\end{equation}
where $\Delta \rho_i^t$ is the amount of contribution of UAV $i$ to coalition $S_k$ in terms of sharing resources to carry out the assigned task $k$ and can be defined as follows:
\begin{equation}
\Delta \rho_i^t =\frac{\Upsilon_k}{\sum_{m\in S_k}f_m}f_i,
\end{equation}
where $\Upsilon_k$ is the sum of the resource requirements of task $T_k$ denoting as $\Upsilon_k=\sum_{j=1}^{N_r}\tau_j^k$ and $f_i$ is the sum of the resource contributions of UAV $i$, defined as $f_i=\sum_{j=1}^{N_r}\frac{r_{u_i}^j}{\tau_j^k}$. As such, the coalition reputation of all involved UAVs in the coalition $S_k$ is computed as follows:
\begin{equation}
P(S_k)=\sum_{i=1}^{N_k}\rho_i^t
\end{equation}

\subsection{Formulation of Multi-objective Optimization}\label{objfun}
To consider all aforementioned optimization criteria including reducing the coalition cost, and increasing the reliability, and reputation of the formed coalitions, Multi-Objective Optimization Problem (MOOP) as a weighted-sum of three objectives is defined. The multi-objective optimization and its required constraints are defined as follows.\\
\begin{equation} \label{multi}
min \qquad O(S_k)= C(S_k)-\eta_1 \ln{R(S_k)} - \eta_2 P(S_k)
\end{equation}
\begin{equation} \label{equ:subj}
subject \quad to \qquad \sum_{i=1}^{N_k}e_{ij}\geq \tau_j^k \quad \forall j= 1,2,3,\ldots,N_r,
 \end{equation}
where $\eta_1$ and $\eta_2$ are weighting parameters to assign the desired importance to each objective and scale them to be in comparative ranges. Constraint (\ref{equ:subj}) refers to the requirement to secure enough resources in the formed coalition to complete the encountered task $T_k$.
In table \ref{tab:notations}, a summary of the notations used throughout this paper is presented. 
\begin{table} 
\caption{Notations}
\label{tab:notations}
\begin{tabular}{|l| p{8cm}|}
\hline
$a_{ij}$  & Travel time of UAV $i$ to task $j$. \\ \hline
$e_{ij}$  & Execution cost of involving resource $j$ of UAV $i$ in a coalition. It is computed per unit time. \\ \hline
$k_{ij}$  & Execution time of involving resource $j$ of UAV $i$ in a coalition. It depends on the task and capability of the UAV. \\ \hline
$\lambda_{ij}$  & Failure rate of involving the resource $j$ of UAV $i$ in a coalition. \\ \hline
$\rho_i$  &  Credit of UAV $i$. \\ \hline
$N_k$  & Number of UAVs in coalition $k$. \\ \hline
$N_r$  & Number of network resources. \\\hline
$N$  & Number of network UAVs. \\\hline
\end{tabular}
\end{table} 

\section{Proposed algorithm}
\label{Sec:proposed}
The objective function described in section \ref{objfun} is a NP-hard problem. Standard approaches such as dynamic programming and exact algorithms involve computational complexity of $O(n^2)$ that could be intractable in large-scale networks. Hence, evolutionary algorithms such as genetic algorithm can be considered as potential options to find feasible solutions of this problem. In this paper, we propose a coalition formation algorithm based on a version of genetic algorithm called Quantum-Inspired Genetic Algorithm (QIGA) to find the solution of the multi-objective problem in (\ref{multi}).

\subsection{Review of Quantum-Inspired Genetic Algorithm (QIGA)}
The idea behind QIG algorithms is to take advantage of both GA and quantum computing mechanisms \cite{han2000genetic}. In quantum computation, the data representation is based on qubit that is the smallest information unit. A qubit is considered as a superposition of two different states $|0 \rangle$ and $|1 \rangle$ that can be denoted as:\\
\begin{equation}
|\psi\rangle=\alpha|0\rangle+\beta|1\rangle,
\end{equation}
where $\alpha$ and $\beta$ are complex numbers such that $|\alpha|^2+|\beta|^2=1$. $|\alpha|^2$ and $|\beta|^2$ are the probability of  amplitudes where the qubit can be at states $|0 \rangle$ and $|1 \rangle$,  respectively. With $m$ qubits, the model can represent $2^m$ independent states. However, when the value of the qubit is measured, it leads to a single state of the quantum state (i.e., $|0 \rangle$ or $|1 \rangle$).
Similar to a standard genetic algorithm, a chromosome's representation is defined as a string of $m$ information units. Thus, a chromosome can be defined as a string of $m$ qubits as:\\
\begin{equation}
\bigg \{ \dbinom{\alpha_0}{\beta_0},\dbinom{\alpha_1}{\beta_1},\dbinom{\alpha_2}{\beta_2}, \ldots, \dbinom{\alpha_{m-1}}{\beta_{m-1}} \bigg \},
\end{equation}
where each pair $(\alpha_i,\beta_i),i=1,2,3,\ldots,m-1$  denotes a gene of the chromosome.

To evaluate a quantum chromosome, a transfer function called \emph{measure} is applied to convert the quantum states to a classical chromosome representation. For example, each pair $(\alpha_i,\beta_i)$ is converted to a value $c_i \in \{0,1\}$ so that the $m$-qubit chromosome may result in a binary string $\{c_1,c_2,c_3,\ldots,c_m\}$. More specifically, each pair becomes $0$ or $1$ using the corresponding qubit probabilities $|\alpha_i|^2$ and $|\beta_i|^2$. To perform this conversion, we use the measure function defined as follow:
\begin{algorithmic} \label{measure}
\Function{measure}{$\alpha$}
\State set $r$  to  a random number between $0$ and $1$;
\If {$r > |\alpha|^2 $}
    \State  \Return $0$;
\Else
\State  \Return $1$;
    \EndIf
\EndFunction
\end{algorithmic}
Each binary string is a possible solution which is evaluated via a problem dependent fitness function. In the following section,  the proposed fitness function is described. 
\subsection{Fitness function}
The fitness function or evaluation function determines how to fit a solution with respect to the constrained optimization problem. To build the fitness function, the negative sign of the objective function $O(S_k )$, defined in (\ref{multi}), is used. In addition, as the solutions which meet all of the constraints get higher fitness value, the solutions which violate some of the constrains should achieve a lower objective value from the fitness function. To prevent potential violations, a penalty function technique that penalizes the solutions according to amount of constraints' mismatches is applied. Considering these facts, the fitness function is proposed as:
\begin{equation}
F(S_k)=-(O(S_k)+g(S_k)),
\end{equation}
where $g(S_k)$ is the penalty function such that if there is no violation, its value will be zero, and positive otherwise. $g(S_k )$ is defined as:
\begin{equation}
g(S_k)= \gamma \times \sum_{i=1}^{N_r}max(0,\tau_j^k-\sum_{i=1}^{N_s}e_{ij}),
\end{equation}
where $\gamma$ is the penalty coefficient that controls the weight of the amount of constraints violated.

\begin{table*}  
\caption{Algorithms' parameters}
\centering{
\label{tab:parameters}
\begin{tabular}{lccccc}
\toprule
\textbf{} & \multicolumn{3}{c}{\textbf{Method}} \\
\cmidrule(lr){2-4}
\textbf{Parameter} & {$Distance-Based$} & {$NSGA-II$} & {$MOQGA$} \\
\midrule
\texttt{Population size} &$NA$ &$200$ &$200$  \\
\texttt{Maximum number of iterations} &$NA$ &$500$ &$500$ \\ 
\texttt{Number of objectives} &$1$ &$3$ &$3$ \\ 
\texttt{Mutation probability} &$NA$ &$10\%$ &$NA$ \\ 
\texttt{Crossover probability} &$NA$ &$90\%$ &$NA$ \\ 
\texttt{Distribution index for crossover} &$NA$ &$20$ &$NA$ \\ 
\texttt{Distribution index for mutation} &$NA$ &$100$ &$NA$ \\ 
        \bottomrule  
NA: Not Available \\ 
\end{tabular}}
\end{table*}

Evolutionary strategies (e.g., crossover and mutation operations) are often considered in GA algorithms to improve their performance. However, as stated in \cite{han2000genetic}, applying the crossover and mutation operators do not significantly improve the performance of the QIGA. Therefore, in the QIGAs, usually a qubit rotation gates strategy is used. Qubit $(\alpha_i,\beta_i)$ of the $m$-qubit chromosome is updated according to the rotation gates in order to get more or less probabilities to states $|0 \rangle$ and $|1 \rangle$. Therefore, at each time step $t$, the update of qubit is performed based on the following rotation gate matrix $L(\theta_i)$:

\begin{equation}
  L(\theta_i)=
  \begin{pmatrix}
    cos(\theta_i) & -sin(\theta_i) \\
    sin(\theta_i) & cos(\theta_i)
  \end{pmatrix}; 
    \begin{pmatrix}
    \alpha_i^t  \\
    \beta_i^t 
  \end{pmatrix}=L(\theta_i) \begin{pmatrix}
    \alpha_i^{t-1}  \\
    \beta_i^{t-1} 
  \end{pmatrix},
  \label{equ:mat}
\end{equation}
where $\theta_i$ is the amount of angle rotation of qubit gate $i$. Algorithm \ref{alg1} presents the pseudo-code for QIGA as described in \cite{han2000genetic}.

\begin{algorithm}
\caption{Quantum-Inspired Genetic Algorithm}
\begin{algorithmic}[1]
\State $t\gets 0$
\State Initialize $Q(t)$ as the population
\State Make $P(t)$ by measuring $Q(t)$
\State Evaluate $P(t)$
\State Store the best solution $b$ among $P(t)$
\Repeat
  \State $t\gets t+1$
  \State Make $P(t)$ by measuring $Q(t-1)$
  \State  Evaluate $P(t)$
  \State Update $Q(t)$ using quantum gates $L(t)$
  \State  Store the best solution $b$ among $P(t)$
\Until {the termination-condition}
\end{algorithmic}
\label{alg1}
\end{algorithm}


\subsection{Multi-UAV Coalition Formation}
To establish a multi-UAV coalition, a leader-follower coalition formation method is followed. 
Initially, the UAVs are uniformly distributed in a search space to look for the PoIs. When a PoI is detected by a UAV, this UAV computes the resource requirements of the detected PoI and serves as a leader to form an optimal coalition. 
After calculating the required resources, the leader UAV calls other UAVs within a  certain distance to join it in forming a coalition. Then, the UAVs with at least one of the required resources can respond to this call by reporting the amount of resources that they are able to contribute to help accomplish the task. It is also assumed that the UAVs are self-interested, meaning that if a UAV receives multiple requests, it will consider joining the coalition which offers the highest benefit. The UAV $i$, $u_i$ measures the value of each request based on travel time to reach the task and the expected cooperative reputation credit received. Thus, its utility value can be defined as:
\begin{equation}
U(u_i,S_k)=\rho_i-\delta a_{ij},
\end{equation}
where $\rho_i$ and $a_i$ are the cooperation credit and travel time of $u_i$ when it attempts to join coalition $k$. $\delta$ is the weight indicating the relative significance of the travel time compared to the expected credit. Algorithm \ref{pseudoMOQGA} shows the pseudo-code for the multi-UAV coalition formation.
\begin{algorithm}
\caption{Multi-UAV coalition formation using the leader-follower method and QIGA}
\label{pseudoMOQGA}
\begin{algorithmic}[1]
\State Search PoIs in the search space
\State Initialize each leader as a singleton coalition committed to a single PoI 
\State $coalition\_members \gets [\,]$
\While{there exists an idle UAV around}
\For{\textbf{all} unsatisfied PoIs so far}
   \State $coal\_mems \gets$ execute MOQIG method regarding algorithm \ref{alg1}
    \State $coalition\_members.append(coal\_mems)$ 
\EndFor
\State Send bids to UAVs as potential followers
\State Calculate the utility values of the followers and receive bid responses
\State Update coalition members of each leader with respect to the bid responses
\EndWhile
\end{algorithmic}
\end{algorithm}


\begin{figure*}
    \centering
    \begin{subfigure}[b]{0.4\textwidth}
         \includegraphics[width=\textwidth]{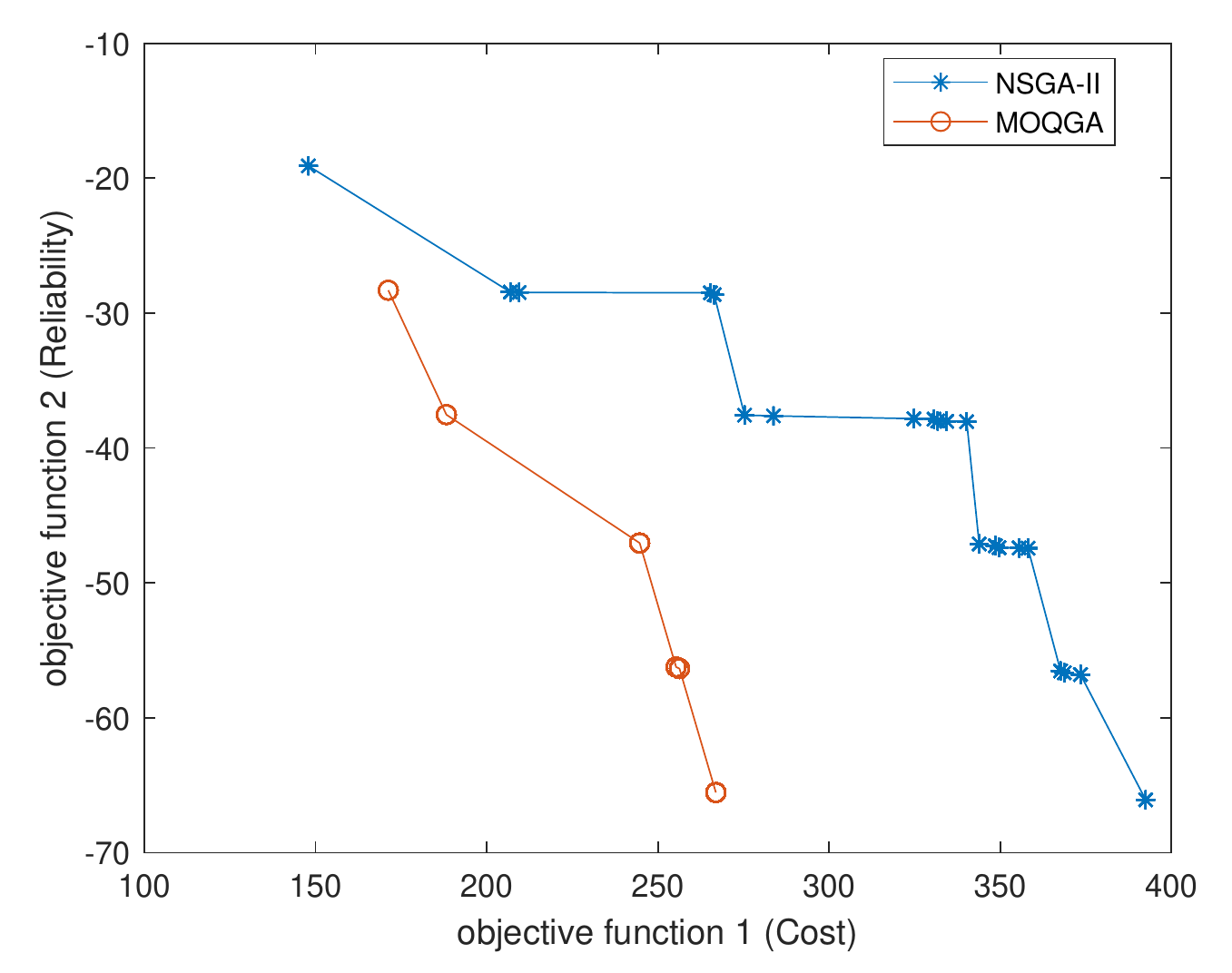}
        \caption{8-2}
        \label{fig:pareto-8-2}
    \end{subfigure}
    ~ 
    \begin{subfigure}[b]{0.4\textwidth}
        \includegraphics[width=\textwidth]{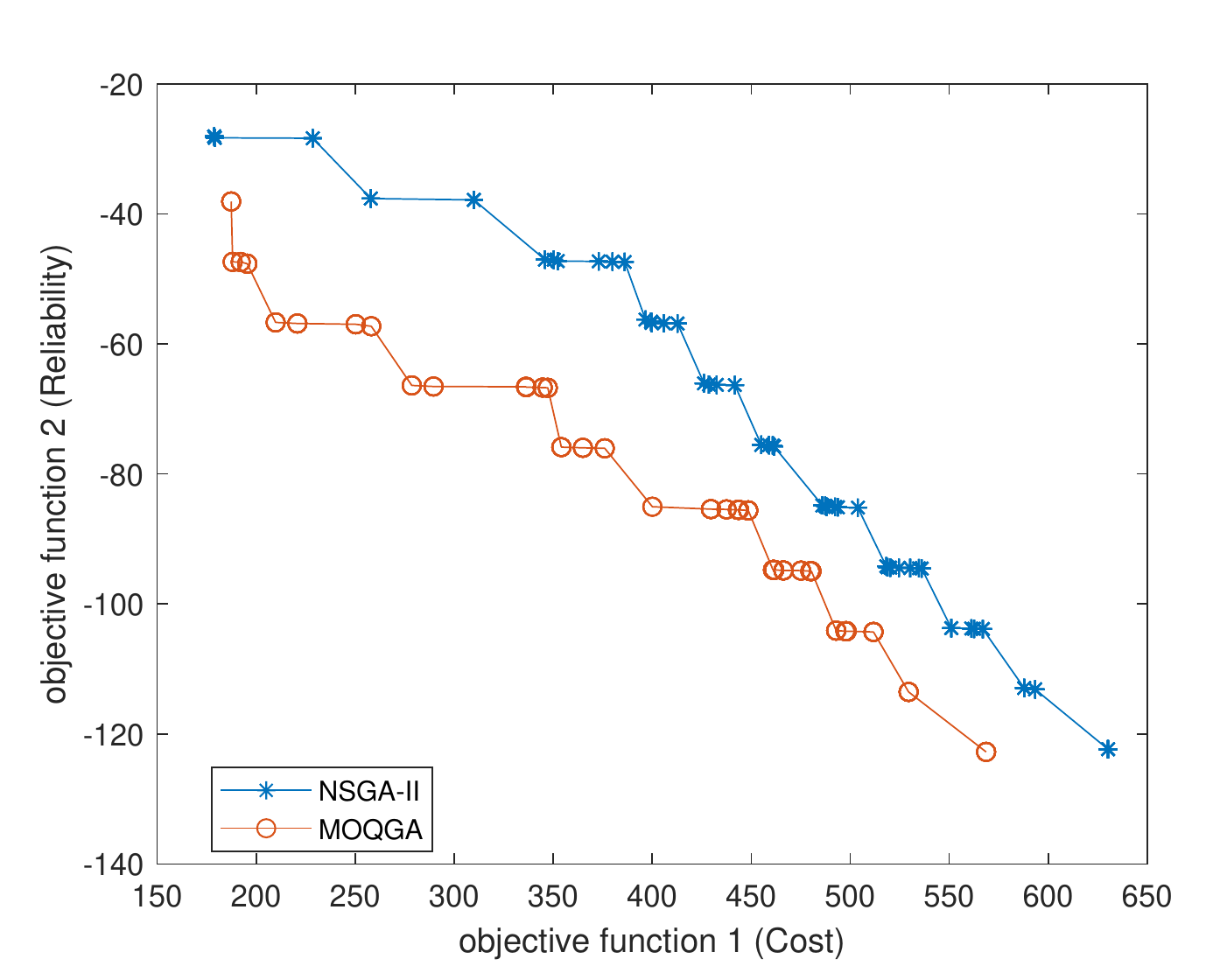}
        \caption{16-4}
        \label{fig:pareto-16-4}
    \end{subfigure}
    ~ 
    
    \begin{subfigure}[b]{0.4\textwidth}
        \includegraphics[width=\textwidth]{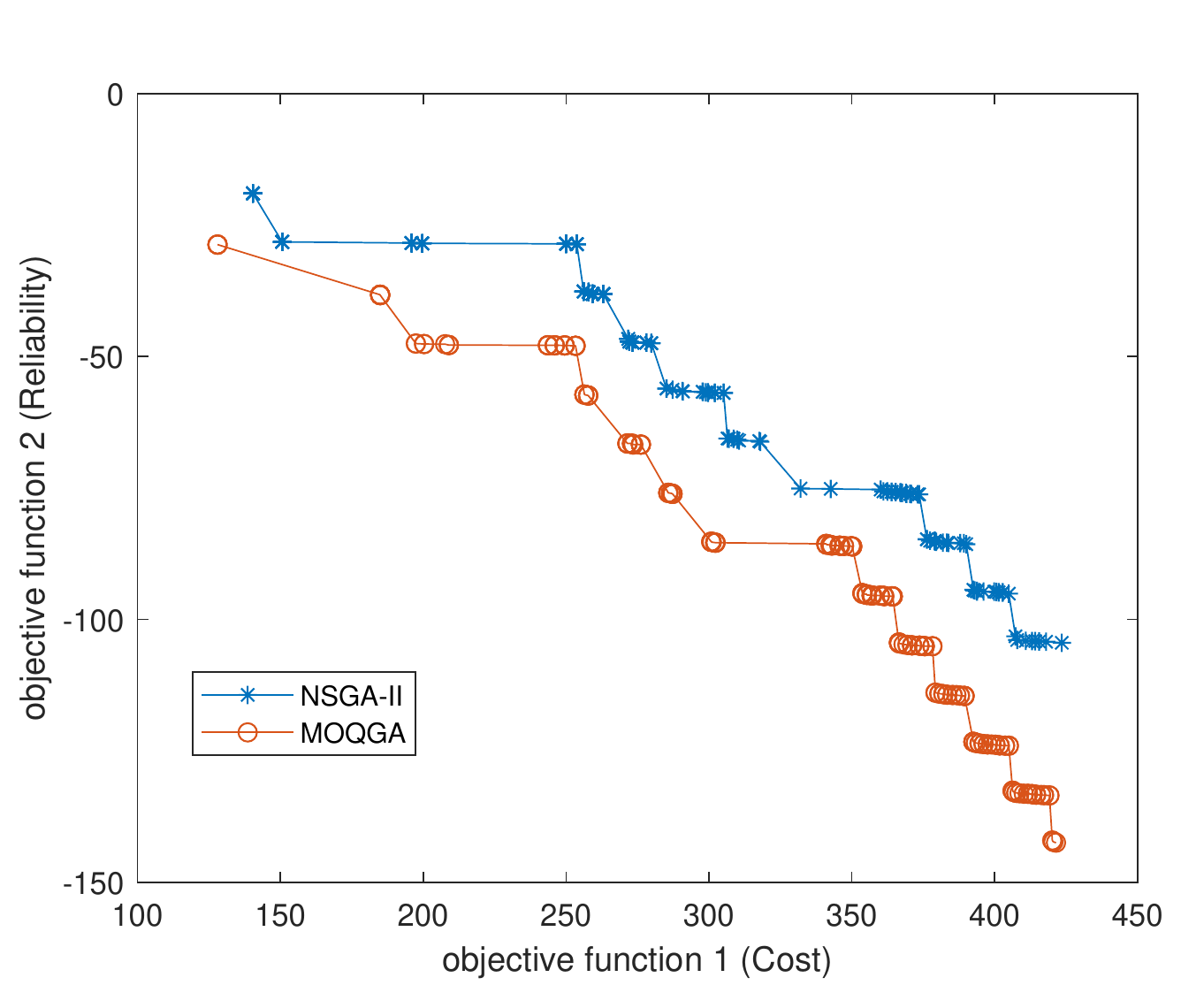}
        \caption{32-8}
        \label{fig:pareto-32-8}
    \end{subfigure}
        \begin{subfigure}[b]{0.4\textwidth}
        \includegraphics[width=\textwidth]{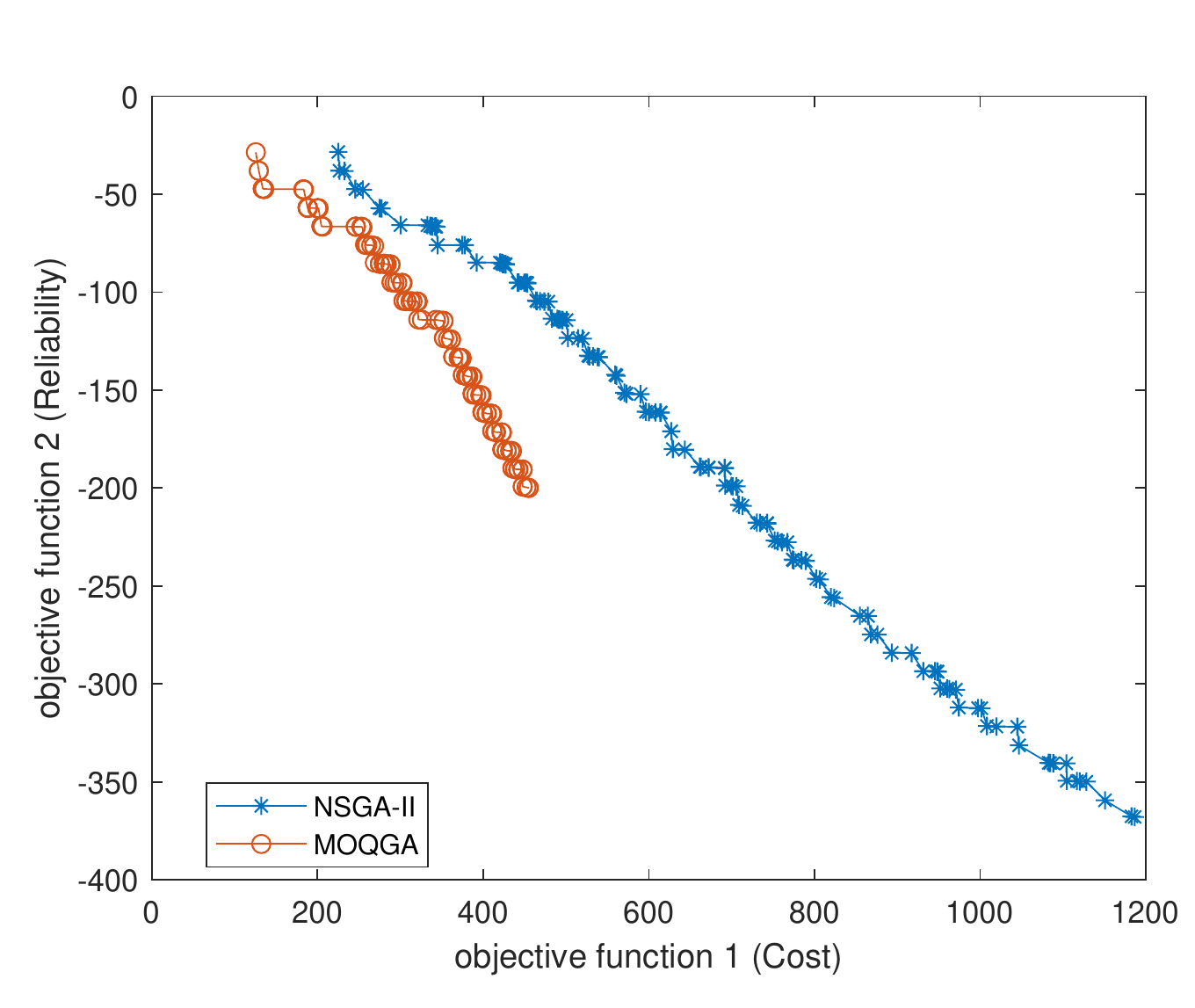}
        \caption{64-16}
        \label{fig:pareto-64-16}
    \end{subfigure}
    \caption{Comparison the qualities of solutions provided by the proposed method (MOQGA) and NSGA-II.}\label{fig:quality}
\end{figure*}

\begin{figure}
\centering

\includegraphics[scale=0.5]{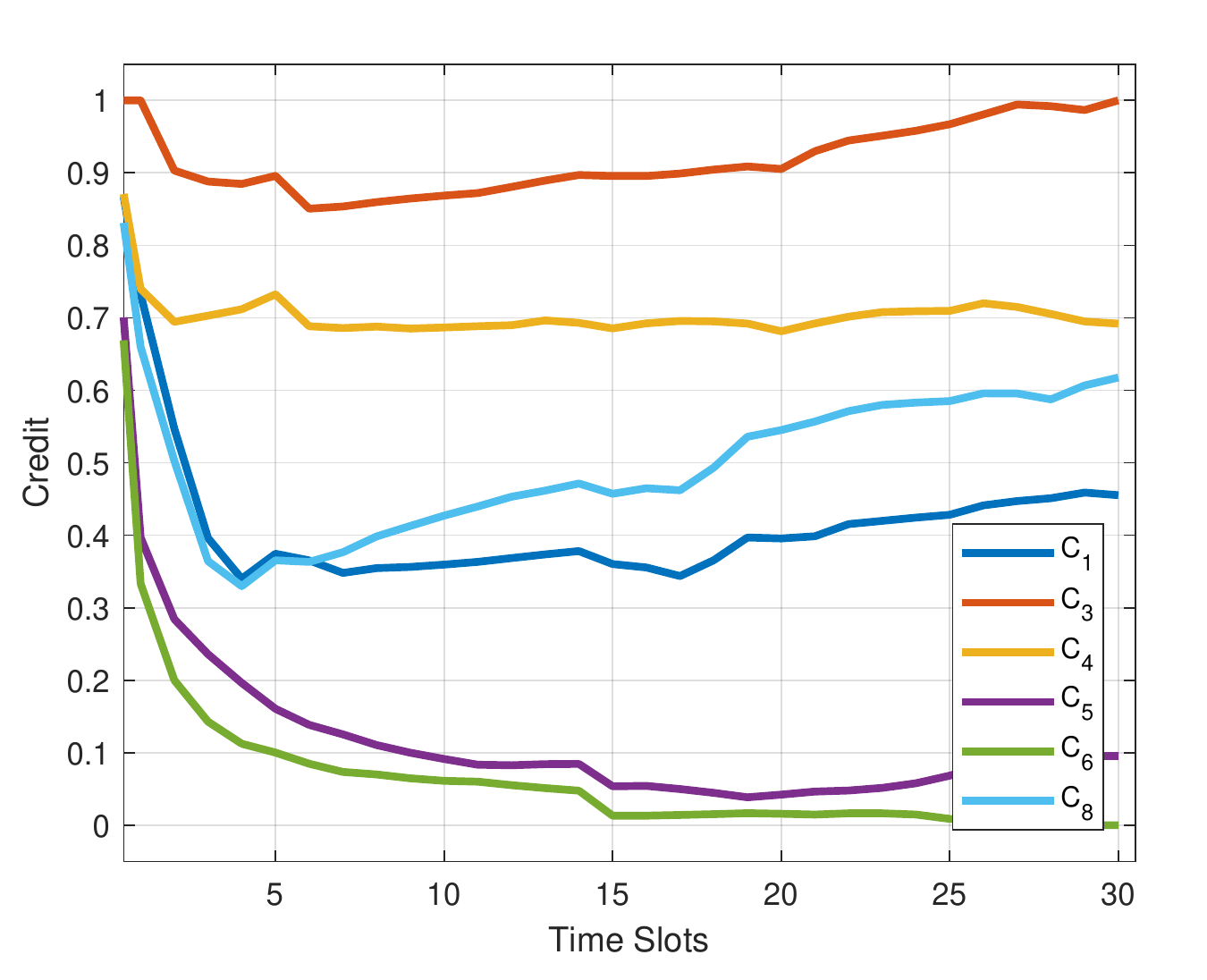}
\caption{Changes in UAVs' cooperative reputation over time, where UAVs $u_5$ and $u_6$ are assumed to be selfish}
\label{fig:8-2-credit}
\end{figure}

\section{Experimental results}
\label{Sec:results}
To evaluate the performance of the proposed QIGA-based coalition formation method, several scenarios with different number of UAVs and PoIs were simulated. It is assumed that the UAVs and PoIs are uniformly distributed across the region and the closest UAV to the PoI is considered to be the one that first detects the PoI and form a coalition (as a coalition leader). It is also assumed that each UAV has five different types of resources. The values for the UAVs' resources are generated with a random uniform distribution and the UAV's resource failure rates are produced randomly in the range $(5\times10^{-5}, 10^{-4})$. Furthermore, the execution times of the identified tasks are computed randomly in the range between $10$ and $20$, depending on the task and the capability of the UAV.
\begin{table*}  
\caption{ Percentage of completed tasks and average of resource violations for different algorithms in 30 missions (with different numbers of UAVs and tasks).}
\centering{
\label{tab:results}
\begin{tabular}{ccccccc}
\toprule
\textbf{} & \multicolumn{6}{c}{\textbf{Method}} \\
\cmidrule(lr){2-7}
 \textbf{} & \multicolumn{2}{c}{$\mathbf{Distance-Based}$} & \multicolumn{2}{c}{$\mathbf{NSGA-II}$} & \multicolumn{2}{c}{$\mathbf{MOQGA}$} \\
 \cmidrule(lr){2-7}
 \textbf{No. of UAVs and tasks} &  Completed tasks & Resource violations & Completed tasks & Resource violations & Completed tasks & Resource violations \\
  \midrule
8-2 &$67\%$ &$1.6$ &$$80\%$$ &$0.43$ &$\underline{\mathbf{90\%}}$ &$\underline{\mathbf{0.3}}$\\
16-4 &$43\%$ &$3.83$ &$$86\%$$ &$0.6$ &$\underline{\mathbf{90\%}}$ &$\underline{\mathbf{0.43}}$\\
32-8 &$39\%$ &$6.93$ &$$91\%$$ &$0.73$ &$\underline{\mathbf{97\%}}$ &$\underline{\mathbf{0.20}}$\\
64-16 &$45\%$ &$11.97$ &$$92\%$$ &$2.03$ &$\underline{\mathbf{95\%}}$ &$\underline{\mathbf{1.23}}$\\
128-24 &$47\%$ &$20.33$ &$$90\%$$ &$4.33$ &$\underline{\mathbf{94\%}}$ &$\underline{\mathbf{2.47}}$\\
\bottomrule             
\end{tabular}}
\end{table*}

\begin{table*} 
\caption{The selected UAVs by leaders in 10 missions, where it is assumed the failure rate of the UAVs $u_5$ and $u_6$ is 90\%.}
\label{tab:reliability}
\centering{
\begin{tabular}{ |c|c|c|c|c|c|} 
\hline
Time slot & Coalition 1  & Satisfied & Coalition 2 & Satisfied  & Unreliable UAVs \\
\hline
$1$ & $L\{u_6\};F\{u_3,u_4,u_8\}$ & Yes &  $L\{u_1\};F\{u_7,u_2\}$ & Yes & \{\} \\ 
\hline
$2$ & $L\{u_2\};F\{u_3,u_8,u_4\}$ & Yes &  $L\{u_1\};F\{u_3,u_7\}$ & Yes & \{\} \\ 
\hline
$3$ & $L\{u_2\};F\{u_3,u_5,u_8\}$ & Yes &  $L\{u_6\};F\{u_4\}$ & Yes & $\{u_5\}$ \\ 
\hline
$4$ & $L\{u_2\};F\{u_4,u_7,u_3\}$ & Yes &  $L\{u_1\};F\{u_5,u_6,u_8\}$ & No & $\{u_5,u_6\}$ \\ 
\hline
$5$ & $L\{u_5\};F\{u_1,u_3,u_4\}$ & Yes &  $L\{u_2\};F\{u_7,u_8\}$ & Yes & \{\} \\ 
\hline
$6$ & $L\{u_1\};F\{u_3,u_4\}$ & Yes &  $L\{u_2\};F\{u_5,u_6,u_8,u_7\}$ & No & $\{u_5,u_6\}$ \\ 
\hline
$7$ & $L\{u_5\};F\{u_2,u_3,u_8\}$ & Yes &  $L\{u_1\};F\{u_4,u_7\}$ & Yes & \{\} \\ 
\hline
$8$ & $L\{u_5\};F\{u_2,u_3,u_6\}$ & Yes &  $L\{u_1\};F\{u_4,u_7\}$ & Yes & $\{u_6\}$ \\ 
\hline
$8$ & $L\{u_5\};F\{u_2,u_6,u_7,u_8\}$ & Yes &  $L\{u_1\};F\{u_3,u_4\}$ & Yes & $\{u_6\}$ \\ 
\hline
$9$ & $L\{u_5\};F\{u_2,u_7,u_8\}$ & No &  $L\{u_1\};F\{u_3,u_4\}$ & Yes & \{\} \\ 
\hline
$10$ & $L\{u_5\};F\{u_1,u_3,u_4\}$ & Yes &  $L\{u_2\};F\{u_7,u_8\}$ & No & \{\} \\ 
\hline
\multicolumn{1}{c}{L: Leader, F: Follower}
\end{tabular} }
\end{table*}
The performance of our proposed algorithm is compared with three well-known algorithms including: i) the distance-based coalition formation method in which the coalitions are formed with the closest UAVs to the leader (i.e., the leader only considers the UAVs in a certain distance of the PoI to be in the coalition and do not evaluate them in terms of cooperative reputation or available resources), ii) the common merge-and-split coalition formation \cite{afghah2017coalition}, and iii) a Non-Dominated Sorting Genetic Algorithm (NSGA-II) which is a fast, elitist and heuristic-based multi-objective algorithm. Table \ref{tab:parameters} shows the corresponding parameters for these algorithms.

Table \ref{tab:results} represents some statistics regarding the completed tasks and resources violations for different algorithms. It demonstrates that the performance of the proposed coalition formation method is quite better than other algorithms addressed here in terms of percentage of completed tasks and average of resource violations. We  also compared the proposed method against the merge-and-split coalition formation algorithm. The percentage of completed tasks for the merge-and-split method in 30 missions for different numbers of UAVs and tasks were between 46\% and 50\%, while, as shown in table \ref{tab:results}, our method could achieve rate of 90\% in task completion.

Figure \ref{fig:quality} compares the qualities of solutions of the MOQGA to NSGA-II algorithm. The plots demonstrate that the performance of the proposed method is better than the NSGA-II algorithm and result in superior quality solutions in terms of lower cost and higher reliability.
Figure \ref{fig:8-2-credit} shows changes in UAV's cooperative reputation over the course of time. It is assumed that UAVs, $u_5$ and $u_6$ are not trustworthy. As seen in the figure, the reputations of these UAVs decrease at each time slot, therefore it is less likely that these UAVs are selected by the leaders over time. 
The impact of reliability in coalition formation is also studied, where a pre-defined failure rate of 90\% is considered for resources of UAVs $u_5$ and $u_6$. Table \ref{tab:reliability} represents the selected UAVs (e.g., followers) by leaders where there are two unreliable UAVs. As shown in the table, the proposed method tries to not select unreliable UAVs in most cases. 
The reason for the selection of unreliable UAVs in some cases is that the problem is a MOOP and the method has to consider other objectives (i.e., cost and reputation) to optimize as well.


\section{Conclusion}
\label{Sec:conclusion}
A leader-follower UAV coalition formation method is proposed to provide a practical solution for distributed task allocation in an unknown environment. Three critical aspects of cost minimization, reliability maximization, and the potential selfish behavior of the UAVs were considered in this coalition formation, and a quantum-inspired genetic algorithm is proposed to find the optimal coalitions with a low level of computational complexity. The proposed approach led to promising results compared to existing solutions with respect to completing a higher number of tasks and minimally overspending the resources.

\section{ACKNOWLEDGMENT OF SUPPORT AND DISCLAIMER}
The authors acknowledge the U.S. Government's support in  the publication of this paper. This material is based upon work funded by AFRL. Any opinions, findings and conclusions or recommendations expressed in this material are those of the author(s) and do not necessarily reflect the views of the US government or AFRL.

\bibliographystyle{IEEEtran}
\bibliography{IEEEtran}

%





\end{document}